\documentclass{ws-p10x7}

\begin{document}

\title{Neutrino Mixing from the CKM matrix in SUSY $SO(10) \times U(2)_{F}$}

\author{Mu-Chun Chen and K.T. Mahanthappa}

\address{Department of Physics, University of Colorado, Boulder, CO 80309-0390,
USA\\E-mail: mu-chun.chen@colorado.edu, ktm@verb.colorado.edu}

\twocolumn[\maketitle\abstract{
We construct a realistic model based on SUSY $SO(10)$ with $U(2)$ flavor
symmetry. 
A set of symmetric mass textures give rise to very good 
predictions; 15 masses and 6 mixing angles are predicted by 11
parameters. Both the vacuum oscillation and LOW solutions are favored
for the solar neutrino problem.
}]

The flavor problem with hierarchical fermion masses and mixing has attracted a
great deal of attention especially since the advent of the atmospheric
neutrino oscillation data from Super-Kamiokande indicating
non-zero neutrino masses. The non-zero neutrino masses give support to the
idea of grand unification based on $SO(10)$ in which all the 16 fermions
(including $\nu_{R}$) can be accommodated in one single spinor
representation. Furthermore, it provides a framework in which seesaw
mechanism arises naturally. 
Naively one expects, for symmetric mass textures, six texture zeros in the 
quark sector. But it has been observed by Ramond, Roberts and Ross\cite{RRR}
that the highest number of texture zeros has to be five, and using
phenomenological analyses, they were able to arrive at five sets of up- and
down-quark mass matrices with five texture zeros. Our analysis with recent
experimental data and using CP conserving real symmetric matrices indicates
that only one set remains viable. The aim of this talk, based on Ref.[2], 
is to present a realistic model based on $SO(10)$ combined with $U(2)$ as
the flavor group, utilizing this set of symmetric mass textures for charged
fermions. We first discuss the viable phenomenology of mass textures
followed by the model which accounts for it, and then the implications
of the model for neutrino mixing are presented. 

The set of up- and down-quark mass matrix combination is given by,
at the GUT scale,
\begin{equation}
\label{eq:Mud}
M_{u} =
\left( \begin{array}{ccc}
0 & 0 & a \\
0 & b & c \\
a & c & 1
\end{array} \right) d, \quad
M_{d}=
\left( \begin{array}{ccc}
0 & e & 0 \\
e & f & 0 \\
0 & 0 & 1
\end{array} \right) h
\end{equation}
with $a \stackrel{<}{\sim} b \ll c \ll 1$, and $e \ll f \ll 1$.
Symmetric mass textures arise naturally if $SO(10)$ breaks down to the SM
group via the left-right symmetric breaking chain $SU(4) \times SU(2)_{L}
\times SU(2)_{R}$. The $SO(10)$ symmetry relates the up-quark to the Dirac
neutrino mass matrices, and the down-quark to the charged lepton mass
matrices. To achieve the Georgi-Jarlskog relations, 
$m_{d} \simeq 3 m_{e}$, $m_{s} \simeq \frac{1}{3} m_{\mu}$, 
$m_{b} \simeq m_{\tau}$, 
a factor of $-3$ is needed
in the $(2,2)$ entry of  the charged lepton mass matrix,  
\begin{equation}
\label{eq:Me}
M_{e}= 
\left( \begin{array}{ccc}
0 & e & 0 \\
e & -3f & 0 \\
0 & 0 & 1
\end{array} \right) h
\end{equation}
This factor of $-3$ can be generated by the $SO(10)$ Clebsch-Gordon 
coefficients through the couplings to the $\overline{126}$ dimesional
representations of Higgses. 
In order to explain the smallness of the neutrino masses, we will adopt
the type I seesaw mechanism which requires both
Dirac and right-handed Majorana mass matrices to be present in the
Lagrangian. 
The Dirac neutrino mass matrix is identical to
the one of the up-quark in the framework of $SO(10)$   
\begin{equation}
\label{eq:Mlr}
M_{\nu_{LR}}=
\left( \begin{array}{ccc}
0 & 0 & a \\
0 & b & c \\
a & c & 1
\end{array} \right) d
\end{equation}
The right-handed neutrino sector is an unknown sector. We find\cite{cm} that if
the right-handed neutrino mass matrix has the same texture as that of the
Dirac  neutrino mass matrix,
\begin{equation}
\label{eq:Mrr}
M_{\nu_{RR}}=
\left( \begin{array}{ccc}
0 & 0 & \delta_{1} \\
0 & \delta_{2} & \delta_{3} \\
\delta_{1} & \delta_{3} & 1
\end{array} \right) M_{R}
\end{equation}
and if the elements $\delta_{i}$ are of the right orders of magnitudes, 
determined by $\delta_{i}=f_{i}(a,b,c,t)$, 
then the resulting effective neutrino mass matrix will take the following form
\begin{equation}
\label{eq:Mll}
M_{\nu_{LL}}=M_{\nu_{LR}}^{T} M_{\nu_{RR}}^{-1} M_{\nu_{LR}}
= \left( 
\begin{array}{ccc}  
0 & 0 & t \\
0 & 1 & 1 \\  
t & 1 & 1 
\end{array} \right) \Lambda
\end{equation}
Note that $M_{\nu_{LL}}$ has the same texture as that of $M_{\nu{LR}}$ and
$M_{\nu_{RR}}$. That is to say, the seesaw mechanism is form invariant.  A
generic feature of mass matricies of the type given in Eq.(\ref{eq:Mll}) 
is that they give rise to 
bi-maximal mixing pattern. After diagonalizing this mass matrix, one can see
immediately that the squared mass difference between $m_{\nu_{1}}^{2}$ and
$m_{\nu_{2}}^{2}$ is of the order of
$O(t^{3})$, while the squared mass difference between $m_{\nu_{2}}^{2}$ and $m_{\nu_{3}}^{2}$ 
is of the order of $O(1)$, in units of $\Lambda$. 
For $t \ll 1$, the phenomenologically favored relation $\Delta
m_{atm}^{2} \gg \Delta m_{\odot}^{2}$ is thus obtained.

The $U(2)$ flavor symmetry\cite{u2} is implemented {\it {\'a} la} the
Froggatt-Nielsen mechanism. 
It simply says that the heaviest matter fields acquire their masses through
tree level interactions with the Higgs fields while masses of lighter matter
fields are produced by higher dimensional interactions involving, in addition
to the regular Higgs fields, exotic vector-like pairs of matter fields and the
so-called flavons (flavor Higgs fields). After integrating out superheavy 
$(\approx M)$ vector-like matter fields, the mass terms of the light matter
fields get suppressed by a factor of $\frac{<\theta>}{M}$, where $<\theta>$ is
the VEVs of the flavons and $M$ is the UV-cutoff of the effective theory above
which the flavor symmetry is exact. 
We assume that the flavor scale is higher than the GUT scale. 
The heaviness of the top quark and to suppress the
SUSY FCNC together suggest that the third family of matter fields transform
as a singlet and the lighter two families of matter fields transform as a
doublet under $U(2)$. In the flavor symmetric limit, only the third family has
non-vanishing Yukawa couplings. $U(2)$ breaks down in two steps:
$U(2) \stackrel{\epsilon M}{\longrightarrow} 
U(1) \stackrel{\epsilon' M}{\longrightarrow}
nothing$, 
where $\epsilon' \ll \epsilon \ll 1$ and $M$ is the flavor scale. 
These small parameters $\epsilon$ and $\epsilon'$ are the ratios of the vacuum expectation
values of the flavon fields to the flavor scale. 
Since $\psi_{3}\psi_{3} \sim 1_{S}$, 
$\psi_{3}\psi_{a} \sim 2$, $\psi_{a}\psi_{b} \sim 2 \otimes 2 = 1_{A} \oplus 3$, 
the only relevant flavon fields are in the 
$A^{ab} \sim 1_{A}$, $\phi^{a} \sim 2$, and 
$S^{ab} \sim 3$ dimensional representations of $U(2)$. 
Because we are confining ourselves to symmetric mass textures, we use only
$\phi^{a}$ and $S^{ab}$. 
In the chosen basis, the VEVs various flavon fields could acquire are given
by 
\begin{equation}
\frac{ \left< \phi \right> }{M} \sim O \left( \begin{array}{c}
\epsilon' \\
\epsilon
\end{array} \right), \quad
\frac{ \left< S^{ab} \right> }{M} \sim O \left( \begin{array}{cc}
\epsilon' & \epsilon' \\
\epsilon' & \epsilon
\end{array} \right)
\end{equation}
Putting everything together, a symmetric mass matrix would have the following built-in
hierarchy given by
\begin{equation}
\left( \begin{array}{ccc}
\epsilon' & \epsilon' & \epsilon' \\
\epsilon' & \epsilon & \epsilon \\
\epsilon' & \epsilon & 1 
\end{array} \right)
\end{equation}
Combining $SO(10)$ with $U(2)$, the most general
superpotential  which repects the symmetry 
one could write down is given schematically by
\begin{equation}
W = H(\psi_{3}
\psi_{3} + \psi_{3} \frac{\phi^{a}}{M} \psi_{a} + \psi_{a}\frac{S^{ab}}{M}
\psi_{b}) 
\end{equation}

The superpotential of our model which generates fermion masses is given by
\begin{equation}
W = W_{D(irac)} + W_{M(ajorana)}
\end{equation}
\begin{eqnarray}
W_{D}=\psi_{3}\psi_{3} T_{1} + \frac{1}{M} \psi_{3} \psi_{a}
\left(T_{2}\phi_{(1)}
+T_{3}\phi_{(2)}\right)
\nonumber\\
+ \frac{1}{M} \psi_{a} \psi_{b} \left(T_{4} + \overline{C}\right) S_{(2)}
+ \frac{1}{M} \psi_{a} \psi_{b} T_{5} S_{(1)}
\nonumber\\
W_{M}=\psi_{3} \psi_{3} \overline{C}_{1} 
+ \frac{1}{M} \psi_{3} \psi_{a} \Phi \overline{C}_{2}
+ \frac{1}{M} \psi_{a} \psi_{b} \Sigma \overline{C}_{2}
\nonumber
\end{eqnarray}
where $T_{i}$'s and $\overline{C}_{i}$'s are the $10$ and $\overline{126}$
dimensional Higgs representations of $SO(10)$ respectively, and $\Phi$ and $\Sigma$ are the 
doublet and triplet of $U(2)$, respectively. 
Detailed quantum number assignments
and the VEVs acquired by various scalar fields are given in Ref.[2]. This
superpotential gives rise to the mass textures given in Eq.(\ref{eq:Mud})-(\ref{eq:Mrr}). 
Various entries of these matrices are given in terms of $\epsilon$,
$\epsilon'$, and ratios of Higgs VEVs. Note that, since we use
$\overline{126}$ dimensional representaions of Higgses to generate the heavy
Majorana neutrino mass terms, R-parity symmetry is preserved at all energies. 

With values of up-quark masses, charged lepton masses and the Cabbibo angle,
the input parameters at the GUT scale are determined\cite{cm}. The charged fermion mass 
predictions of our model at $M_{Z}$ which are summerized in 
Table[\ref{table:mass}] including 2-loop RGE effects are in good agreements with the
experimental values\cite{cm}. The CKM matrix is predicted to be
\begin{equation}
\left(\begin{array}{ccc}
0.975 & 0.222 & 0.00354 \\
0.222 & 0.975 & 0.0367 \\
0.00474 & 0.0368 & 0.999
\end{array} \right)
\end{equation}
In the neutrino sector,
the VO solution to the solar neutrino problem is obtained with 
$(\delta_{1}, \delta_{2}, \delta_{3}, M_{R}) = (0.00116,3.32\times10^{-5},
0.0152,1.32\times10^{14}GeV)$. The atmospheric and solar squared mass
differences are predicted to be 
$\Delta m_{23}^{2} = 3.11 \times 10^{-3} eV^{2}$ and 
$\Delta m_{12}^{2} = 2.87 \times 10^{-10} eV^{2}$; the mixing angles are given
by 
$\sin^{2} 2\theta_{atm}=0.999$, and 
$\sin^{2} 2\theta_{\odot}= 0.991$. 
$|U_{e \nu_{3}}|$ is predicted to be $0.0527$ which is below the
upper bound $0.16$ by the CHOOZ experiment. 
We can also have the LOW solution
with $(\delta_{1},\delta_{2},\delta_{3},M_{R})=(0.00115,2.35\times10^{-4},
0.0168,1.62\times 10^{13}GeV)$. In this case, 
$\Delta m_{23}^{2} = 3.97 \times 10^{-3} eV^{2}$, and 
$\Delta m_{12}^{2} = 1.30 \times 10^{-7} eV^{2}$. The mixing angles are given
by 
$\sin^{2} 2\theta_{atm}=0.999$, and 
$\sin^{2} 2\theta_{\odot}=0.990$. 
$|U_{e \nu_{3}}|$ is predicted to be $0.0743$.
It is possible to have the LAMSW
solution with
$(\delta_{1},\delta_{2},\delta_{3},M_{R})=(0.00108,9.87\times10^{-5},
0.0224,2.42\times10^{12}GeV)$. These parameters predict 
$\Delta m_{23}^{2} = 9.85 \times 10^{-3} eV^{2}$, and 
$\Delta m_{12}^{2} = 2.75 \times 10^{-5} eV^{2}$. The mixing angles are 
$\sin^{2} 2\theta_{atm}=1.00$, and 
$\sin^{2} 2\theta_{\odot}=0.985$. However, $|U_{e\nu_{3}}|$ is predicted to
be $0.158$, right at the experimental upper bound. We note that a 
$|U_{e\nu_{3}}|$ value of less than $0.158$ would lead to
$\Delta m_{23}^{2} > 10^{-2} eV^{2}$ 
leading to the elimination of the LAMSW 
solution in our model. This is a characteristic of the LAMSW solution with 
$\Delta m_{12}^{2} \geq 10^{-5} eV^{2}$. 

Other aspects of our model including the proton stability and symmetry 
breaking are under investigation.

\begin{table}
\footnotesize
\begin{tabular}{| l c | l c l |}
\hline
 & & data at $M_{z}$
 & & predictions  \\ 
\hline
$m_{u}$
& & $2.33^{+0.42}_{-0.45}MeV$ 
& & $1.917 MeV$\\
$m_{c}$ 
 & & $677^{+56}_{-61}MeV$  
 & & $738.7 MeV$\\
$m_{t}$ 
 & & $181^{+}_{-}13GeV$ 
 & & $184.3 MeV$\\
$\frac{m_{d}}{m_{s}}$  
& & $17 \sim 25$  
& & $22.5$\\
$m_{s}$ 
& & $93.4^{+11.8}_{-13.0}MeV$  
& & $83.15 GeV$\\
$m_{b}$ 
& & $3.00^{+}_{-}0.11GeV$  
& & $3.0141 GeV$\\
$m_{e}$  
& & $0.486847MeV$  
& & $0.486 MeV$\\
$m_{\mu}$  
& & $102.75MeV$ 
& & $102.8 MeV$\\
$m_{\tau}$  
& & $1.7467 GeV$  
& & $1.744 GeV$ \\
\hline
\end{tabular}
\caption{\linespread{0.9} \scriptsize Predictions and values extrapolated
from experimental data at $M_{Z}$ for charged fermion masses.}    
\label{table:mass}
\end{table}

\section*{Acknowledgments}
This work was supported in part by the US DoE Grant No. DE FG03-05ER40894.


\begin{thebibliography}{99}

\bibitem{RRR}
P. Ramond, R. Roberts and G. Ross, {\it Nucl. Phys.} B {\bf 406}, 19 (1993).

\bibitem{cm}
M.C. Chen and K.T. Mahanthappa, hep-ph/0005292, to appear in {\it Phys. Rev.}
D; this paper contains details and other relevant references.

\bibitem{u2}
R. Barbieri, L.J. Hall, S. Raby and A. Romanino, {\it Nucl. Phys.} 
B {\bf 493}, 3 (1997).

\end{thebibliography}
\end{document}